\documentclass{article}

\usepackage{arxiv}

\usepackage[utf8]{inputenc} 
\usepackage[T1]{fontenc}    
\usepackage{hyperref}       
\usepackage{url}            
\usepackage{booktabs}       
\usepackage{amsfonts}       
\usepackage{nicefrac}       
\usepackage{microtype}      %
\usepackage{enumitem}

\usepackage{natbib}
\usepackage{graphicx}

\title{Drift of pancake ice floes in the winter Antarctic marginal ice zone during polar cyclones}

\author{
  Alberto Alberello\thanks{\texttt{alberto.alberello@outlook.com}} \\
  University of Adelaide\\
  5005, Adelaide, Australia\\
   \And
  Luke Bennetts\\
  University of Adelaide\\
  5005, Adelaide, Australia\\
     \And
  Petra Heil\\
  Australian Antarctic Division \& ACE--CRC\\
  7001, Hobart, Australia\\
       \And
  Clare Eayrs\\
  New York University Abu Dhabi\\
  Abu Dhabi, United Arab Emirates\\
         \And
  Marcello Vichi\\
  University of Cape Town\\
  Rondenbosch, 7701, South Africa\\
           \And
  Keith MacHutchon\\
  University of Cape Town\\
  Rondenbosch, 7701, South Africa\\
           \And
  Miguel Onorato\\
  Universit\`{a} di Torino \& IFNF\\
  Torino, 10125, Italy\\
     \And
  Alessandro Toffoli\\
  The University of Melbourne\\
  3010, Parkville, Australia\\
}

\begin{document}
\maketitle

\begin{abstract}
High temporal resolution in--situ measurements of pancake ice drift are presented, from a pair of buoys deployed on floes in the Antarctic marginal ice zone during the winter sea ice expansion, over nine days in which the region was impacted by four polar cyclones.
Concomitant measurements of wave-in-ice activity from the buoys is used to infer that pancake ice conditions were maintained over at least the first seven days.
Analysis of the data shows: (i)~unprecedentedly fast drift speeds in the Southern Ocean; 
(ii)~high correlation of drift velocities with the surface wind velocities, indicating absence of internal ice stresses $>$100\,km in from the edge in 100\% remotely sensed ice concentration; 
and (iii)~presence of a strong inertial signature with a 13\,h period. 
A Langrangian free drift model is developed, including a term for geostrophic currents that reproduces the 13\,h period signature in the ice motion.
The calibrated model is shown to provide accurate predictions of the ice drift for up to 2\,days, and the calibrated parameters provide estimates of wind and ocean drag for pancake floes under storm conditions. 
\end{abstract}


\section{Introduction}

Sea ice extent modulates energy, mass and momentum exchanges between the ocean and atmosphere, thereby playing a pivotal role in the global climate system \citep{mcphee1987jgr,notz2012review,vihma2014acp}.
During the winter sea ice advance around Antarctica, pancake ice floes---small, roughly circular floes that form in wavy conditions---represent most of the sea ice mass budget \citep{wadhams2018jgr}.
Dynamics and thermodynamics of pancake floes dominate the evolution of the Antarctic marginal ice zone \citep{doble2003pancake,doble2006jgr,roach2018bjgr}, and also the emerging Arctic marginal ice zone \citep{pedersen2004jgr,wadhams2018jgr,roach2018jgr,smith2018jgr}, i.e.~the 5--100\,km wide outer ice belt, where atmosphere--ocean--sea ice interactions are most intense \citep{wadhams1986book,strong2017jaot}.
Contemporary numerical models struggle to predict the spatial variability of advance/retreat of sea ice around Antarctica \citep{hobbs2015jcli,hobbs2016gpc,kwok2017elementa,roach2018cryo}, resulting in strong biases in ocean--atmosphere heat fluxes and salt input to the ocean \citep{doble2009jgr}.

Except for few sectors around Antarctica, trends in sea ice duration and extent are dominated by storms rather than large atmospheric modes \citep{matear2015natcom,kwok2017elementa,schroeter2017tc}.
\citet{vichi2019grl} have shown that intense winter polar cyclones continuously reshape the edge of the Antarctic marginal ice zone by advecting warm air on the sea ice and forcing ice drift.  
This generates strong coupling between thermodynamics and dynamics \citep{stevens2011annals}. 
Strong coupling also exists in the Arctic, where storms have been shown to reverse the winter sea ice advance by melting \citep{smith2018jgr} and drifting \citep{smith2018jgr,lund2018jgr} newly formed pancakes.
Moreover, intense storm events cause rapid ice drift that enhances mixing and deepens the surface mixed layer, thus promoting heat exchanges with the water sublayers \citep{ackley2015aog,zippel2016elementa,castellani2018om}.
Knowledge of the dynamical response of pancake ice floes to the frequent and intense storm events that impact the winter Antarctic marginal ice zone is required to model the evolution of the marginal ice zone and improve climate predictions \citep{schroeter2017tc,barthlemy2018cd}, particularly now that prognostic floe size information is being included in models \citep{bennetts2017cryo,roach2018bjgr}.

For over a century, atmospheric drag has been identified as the main driver of ice drift \citep{nansen1902,shackleton1920}; a rule-of-thumb indicates that the wind factor (the ice to wind speed ratio) is 2\% \citep{thorndike1982jgr,lepparanta2011book}.
In the Arctic marginal ice zone, \citet{wilkinson2003jgr} calculated an average wind factor of 2.7\% for pancake ice, and noted a correlation with the ice concentration ($i_{c}$), with a larger wind factor of 3.9\% towards the ice edge where $i_c<25\%$, and a smaller value of 2.2\% towards the interior of the marginal ice zone where $i_c>75\%$.
For the Antarctic, \citet{doble2006jgr} calculated a wind factor 3--3.5\% in pancake ice conditions.
\citet{doble2006jgr} used a far shorter sampling rate of 0.33\,h than the 24\,h rate used by \citet{wilkinson2003jgr}, possibly resulting in the larger wind factor (see \S\ref{sec:discuss}).
More recently, in the Arctic and for a low ice concentration ($i_c=33\%$), \citet{lund2018jgr} reported wind factors up to 5\% for pancake ice, but defined for wind at 17\,m height, as opposed to the standard 10\,m height. 
They showed low correlation with the wind forcing, and suggested that currents (not measured) contribute significantly to sea ice drift.
The wind factors calculated by \citet{wilkinson2003jgr}, \citet{doble2006jgr}, \citet{lund2018jgr} and others, do not separate out the effect of currents and Coriolis---they assume is wind is the only forcing. 
As a result,  wind stresses are likely to be underestimated \citep{lepparanta2011book}.
The Nansen number, i.e.~the ratio between ice drift and wind speed for an ocean at rest, explicitly indicates the role of wind stresses only \citep{lepparanta2011book}, but, to the best of our knowledge, the Nansen number has not previously been reported for pancake ice.

In sophisticated contemporary models, sea ice drift is governed by a general horizontal momentum equation in which wind stresses act together with other external stresses (ocean drag, Coriolis forcing, waves and ocean tilt as external forcing), and with a rheology term used to model internal stresses \citep{heil2002jpo,lepparanta2011book}.
Granular rheologies have been developed for the marginal ice zone \citep{shen1987jgr,feltham2005ptra}, in which internal stresses are generated by floe--floe collisions, and the magnitude of the internal stresses depends on concentration of the floes and their granular temperature (a measure of the turbulent kinetic energy of the floes).  
However, in low ice concentration internal stresses are small, and the rheology term typically neglected \citep{hunke2010jofg,herman2012cejp}.
Modelled wind-induced stresses are defined by a standard drag formulation \citep{martinson1990jgr}, i.e.~proportional to air--sea ice drag coefficient---usually larger close to the ice edge because of an increase in surface roughness \citep{johannessen1983jgr}---and the relative velocity between wind and ice.
Similarly, ocean--induced stresses are defined by water--sea ice drag coefficient and ocean currents, seldom available in ice covered regions \citep{nakayama2012jpo}.
Scarcity of in--situ observation (conducted for different seasons, regions and ice types) and heterogeneity of ice conditions, particularly in the highly dynamical marginal ice zone \citep{doble2009jgr}, has led to a wide range of sea ice drag coefficients \citep{lepparanta2011book}, undermining predictive capabilities.

We report and analyse a new set of pancake ice drift measurements during the winter expansion of the Antarctic marginal ice zone, and during intense storm conditions that reshaped the edge of the marginal ice zone at synoptic scales \citep{vichi2019grl}.
In 100\% ice concentration, $\approx60$\% pancake floes and $\approx40$\% interstitial frazil ice \citep{alberello2019cryo}, 
we report the fasted ice drift recorded in the Southern Ocean.
We develop a Lagrangian free-drift model, based on the general sea ice horizontal momentum equation,
and quantify the reciprocal effect of winds and currents on pancake ice drift by providing the Nansen number and the derived current. 

\section{Field experiment and prevailing conditions}\label{env}

\begin{figure}[htbp]
\centering
\includegraphics[clip,trim=0 0 0 27cm,width=1\linewidth]{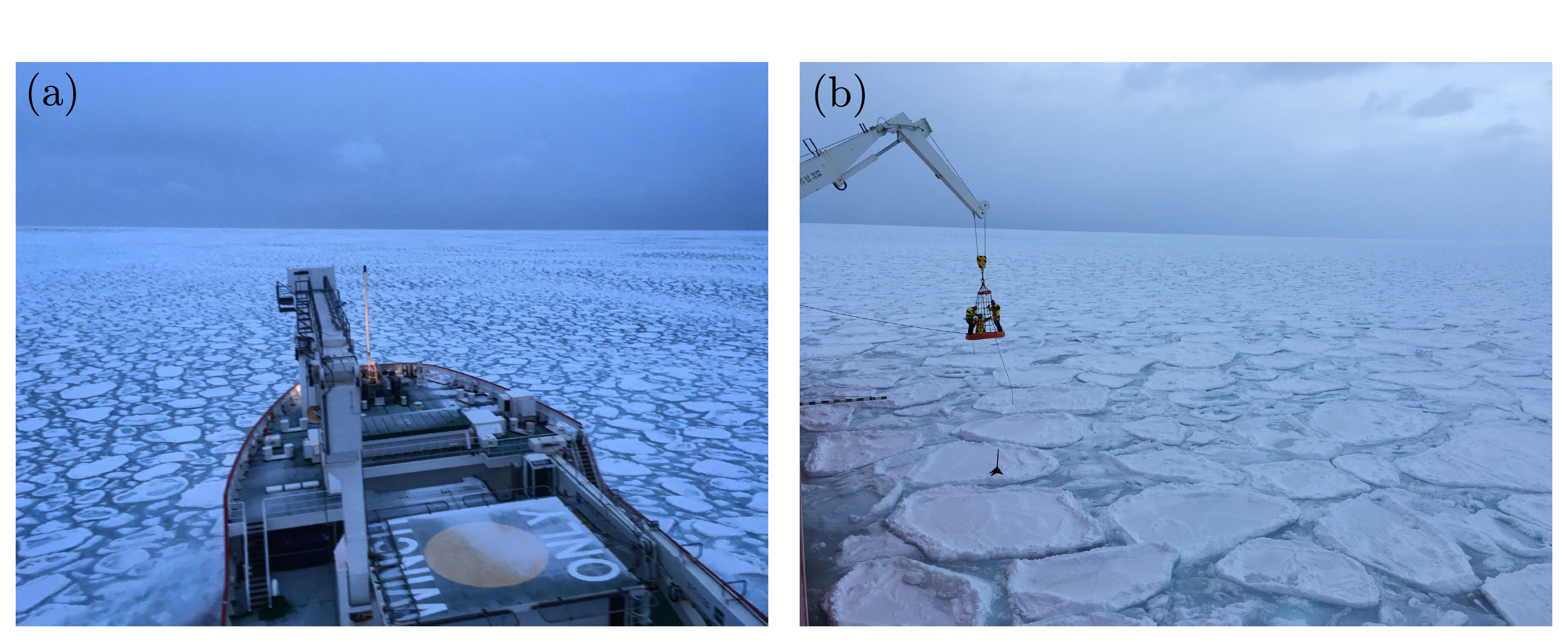}
\caption{(a)~Ice conditions on the 4th~of~July~2017, and (b)~deployment of the buoys on pancake floes using the ship crane.}
\label{f1}
\end{figure}

The instruments were deployed during a winter voyage to the Antarctic marginal ice zone by the icebreaker SA~Agulhas~II (Fig.~\ref{f1}a). 
The voyage departed from Cape Town, South Africa, on the 1st~of~July along the WOCE~I06 transect and reached the marginal ice zone on the 4th~of~July at $62.5^{\circ}$S and $30^{\circ}$E, at which time a polar cyclone was crossing the ice edge.

At midday on the 4th~of~July, a pair of waves-in-ice observation systems \citep{kohout2015annals}, hereafter simply referred to as buoys, were deployed on separate pancake ice floes (Fig.~\ref{f1}b) at 62.8$^\circ$S and 29.8$^\circ$E; they were $\approx$100\,km from the ice edge and $\approx$1\,km apart.  
The buoys are expendable devices that record position and wave spectral characteristics.
One of the buoys, B1, recorded data continuously  at a sampling rate of 15\,mins for almost 9 days---8 days and 18\,h, from 12:00 on the 4th~of~July until 06:00 on the 13th~of~July---until signal was lost (most likely due to the battery running out). The other buoy, B2, recorded at 15\,mins for the first 6 days from deployment, after which, to save battery life, the sampling rate was reduced to 2\,h, which allowed it to record data for 3 weeks. 
In this study, we only consider the period over the 9 days in which both buoys were operational to allow analysis of the buoys' relative motion and ice internal stresses.

Sustained winds over the open ocean, up to 33\,m\,s$^{-1}$ according to the on-board met-station, generated large waves in the open ocean, with significant wave height up to 14\,m and peak period $\approx12$\,s
according to the
ERA5 reanalysis data \citep{era5}, and propagating towards the ice edge. 
The buoys indicated the wave field maintained $\approx50\%$ of its energy after 100\,km of propagation into the marginal ice zone.
Sea ice concentration was $i_c=100$\%, as sourced from AMSR2 \citep{beitsch2014remotesensing} and confirmed by ASPeCt observations \citep{dejong2018sicw}. 
Deck observations (see Fig.~\ref{f1}) and automatic camera measurements revealed the marginal ice zone was an unconsolidated mixture of pancake ice floes covering $\approx$\,60\% of the surface and of characteristic diameter 3.2\,m \citep{alberello2019cryo}, and interstitial frazil ice. In--situ observations in the marginal ice zone lasted $\approx$\,24\,h, after which the ship headed back to Cape Town.

\begin{figure}[htbp]
\centering
\includegraphics[width=1\linewidth]{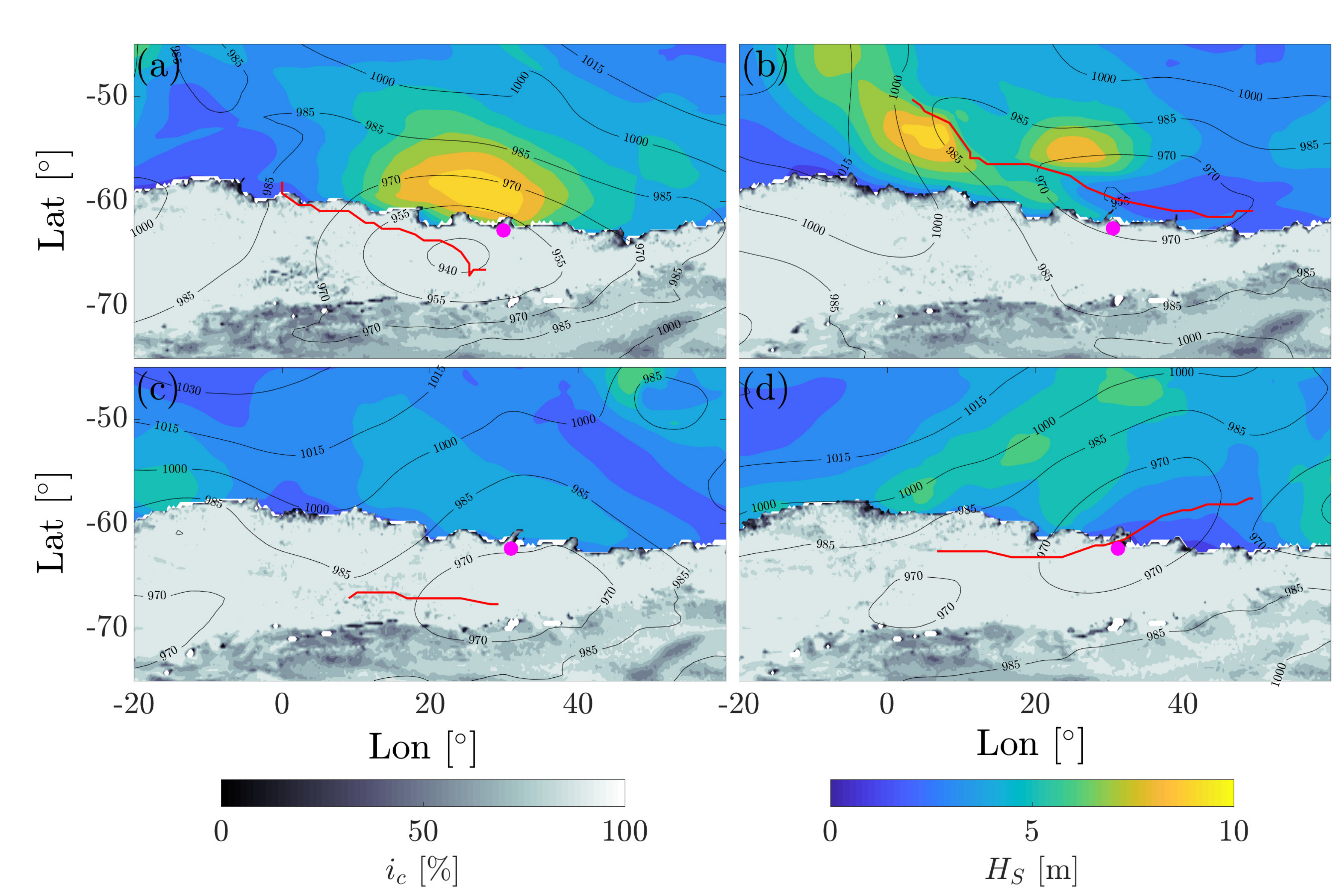}
\caption{Environmental conditions when the polar cyclones were close to the buoys: (a)~at 15:00 on 4th~of~July; (b)~at 18:00 on 7th~of~July; (c)~at 21:00 on 10th~of~July; (d)~at 15:00 on 12th~of~July. The cyclone tracks are shown in red, and the buoy B1 with the magenta circle. The shadings show the AMSR2 ice concentration, and the ERA5 significant wave height. The contour lines (in black) denote the ERA5 isobars in hPa.}
\label{f2}
\end{figure}

Environmental conditions were retrieved from satellite data and reanalysis products over the 9 days both buoys returned data. AMSR2 \citep{spreen2008jgr} provided ice concentration at 3.125\,km spatial resolution as daily mosaics, averaged over two swaths within 24\,h.
ERA5 reanalysis \citep{era5} was used to retrieve surface wind velocities (at 10\,m height) at 0.25$^\circ$ spatial resolution and hourly frequency. ERA5 also provides wave properties, but these are only available where the ice concentration is below 30\% \citep{doble2013om}.

ERA5 reanalysis shows that another three cyclones, albeit less intense, impacted the marginal ice zone  surrounding of the buoys over the 9 days.   
Figs.~\ref{f2}a--d show the tracks of the four polar cyclones overlaid on the significant wave height (in open water) and the ice concentration.
The cyclogenesis of the first and most intense polar cyclone took place over open water and its cyclolysis over the marginal ice zone south-east of the buoys (see track in Fig.~\ref{f2}a).
The second polar cyclone skirted the ice edge, travelling over open water to the north of the buoys (Fig.~\ref{f2}b).
The other two had cyclogenesis over the marginal ice zone. The third cyclone (Fig.~\ref{f2}c) was short lived and its cyclolysis was south of the buoys in the marginal ice zone. The last cyclone transited to the north-west of the buoys before progressing over open water (Fig.~\ref{f2}d).
All observed polar cyclones strongly affected the evolution of the edge of the marginal ice zone at synoptic scale \citep{vichi2019grl}: the asymmetric cyclonic structure transports moist warm air over the sea ice while the opposite side drags ice toward the open ocean.
Concurrently, strong winds associated with polar cyclones generated large waves (larger during the first two polar cyclones that developed over open water) that impacted the edge of the marginal ice zone.

\begin{figure}[htbp]
\centering
\includegraphics[width=1\linewidth]{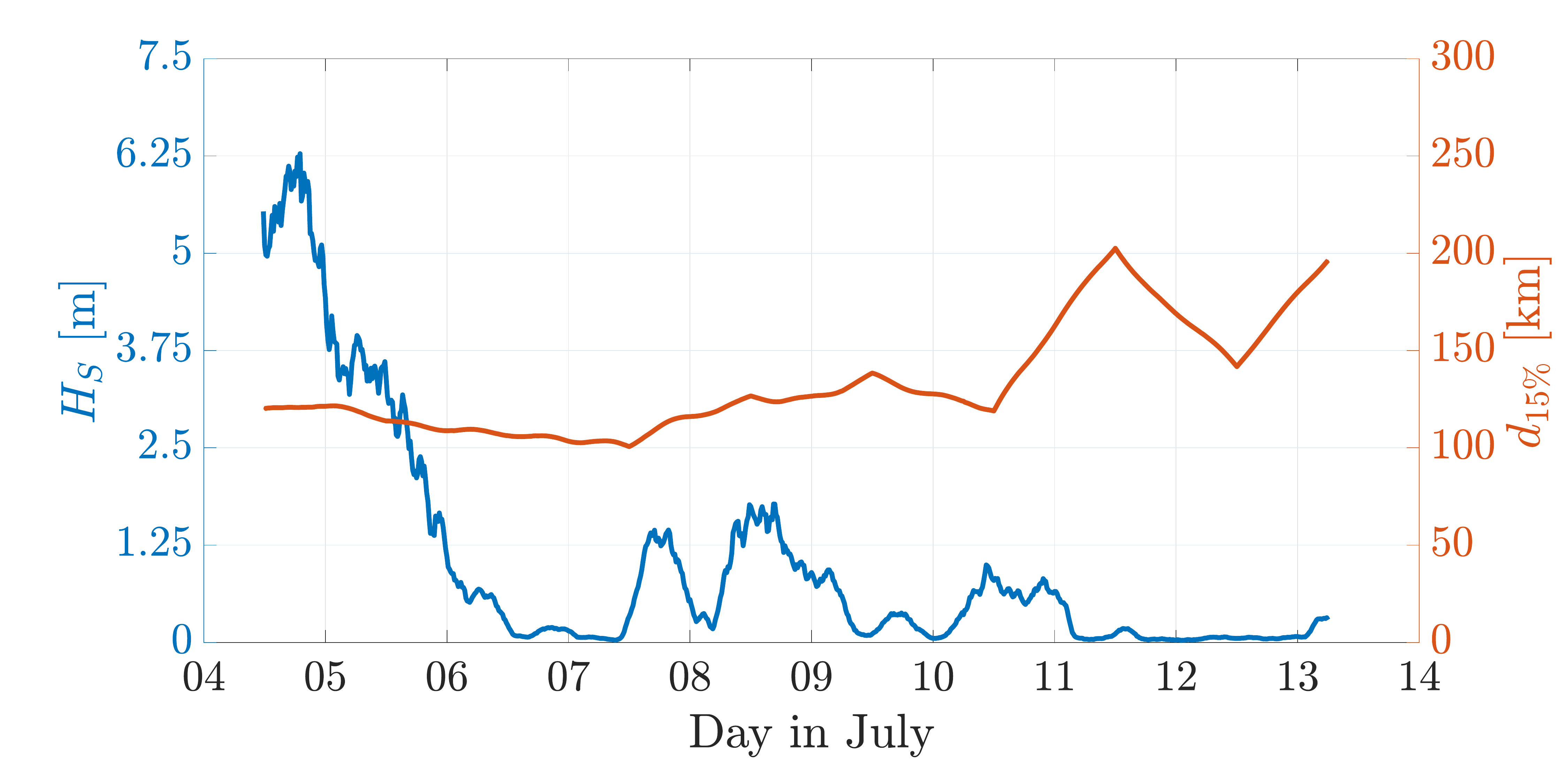}
\caption{Significant wave height from buoy~B1 (left axis; blue), 
distance between buoy~B1 and the ice edge defined by $15\%$ concentration (right axis; orange).}
\label{f3}
\end{figure}

Fig.~\ref{f3} shows wave-in-ice intensity measured by buoy~B1, the peak period was 15--20\,s. 
Peaks in wave activity are associated with the transit of cyclones and they show high correlation with the open water wave height \citep{vichi2019grl}.
Intense wave--in--ice activity after deployment suggests that the marginal ice zone was comprised of pancake floes, at least until the 11th~of~July, when waves ceased.
Fig.~\ref{f3} also shows the distance between buoy~B1 and the ice edge, which is defined as the daily mean position of the AMSR2 15$\%$ ice concentration in the sector 29$^\circ$--33$^\circ$E and the buoy, and denoted $d_{15\%}$.
The buoys are 100--200\,km from the ice edge, noting that sector averaging smears ice-edge features and so the distance must be interpreted with care.

\section{Drift measurements and analysis}

\subsection{Buoy drift}

\begin{figure}[htbp]
\centering
\includegraphics[width=1\linewidth]{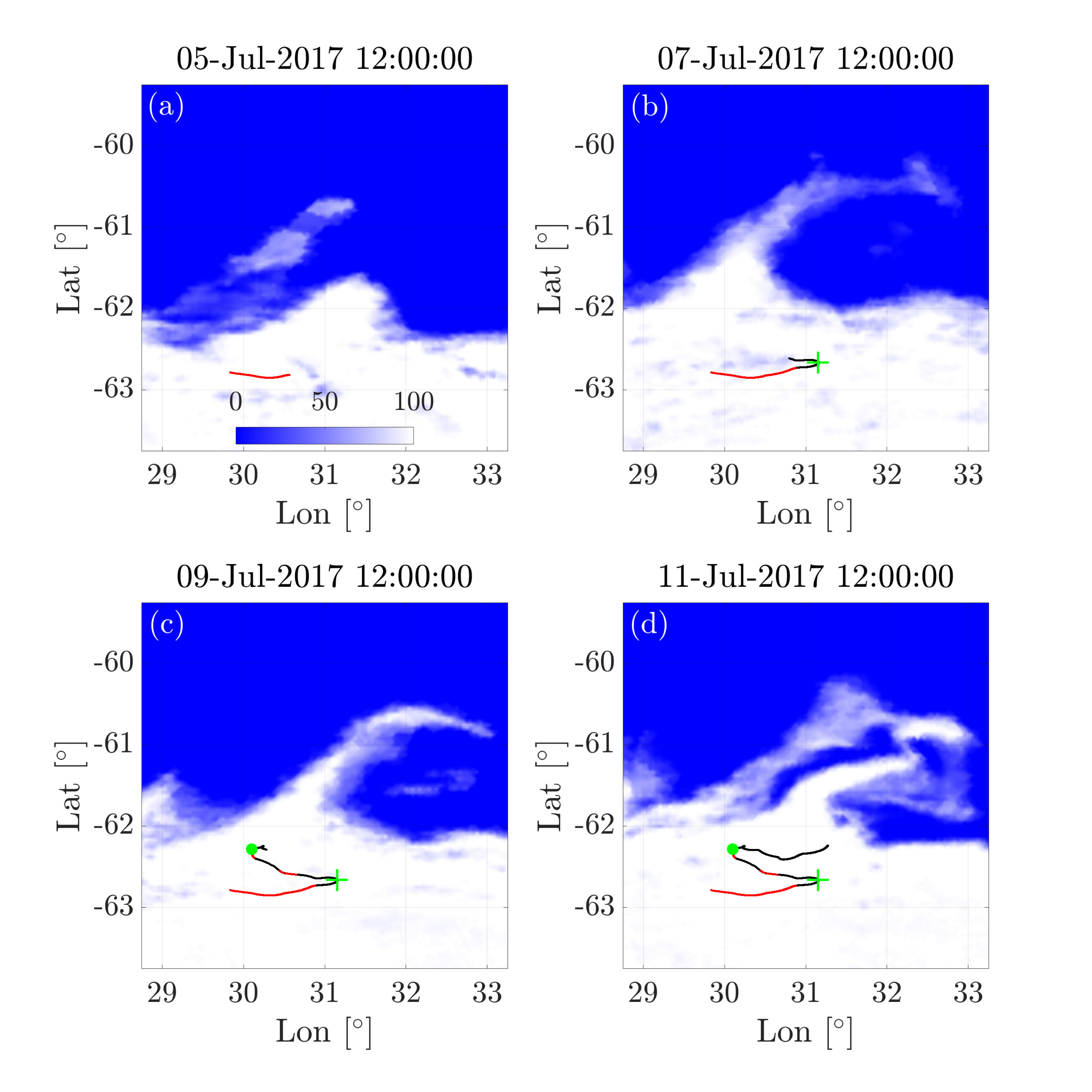}
\caption{Buoy~B1 track over 9 days following deployment on the 4 July 2017, superimposed on ice concentration. Intervals intense wave-in-ice activity ($H_S>1.25$\,m) are highlighted (red). The green cross and the green dot indicate the time of transition between phases}
\label{f4}
\end{figure}

Fig.~\ref{f4} shows the track of buoy~B1 from deployment superimposed on the AMSR2 ice concentration, where each subplot is two days apart. 
(At the scale shown, the track of buoy~B2 would overlap the buoy~B1 track.)
Over the 9 days we identified three distinct phases of ice movement:
\begin{enumerate}[label=\roman*.]
    \item over the first 2 days, and driven by the first cyclone during which the wind speed reached $\approx$15\,m\,s$^{-1}$ (to the east), the drift was predominantly eastward, initially with a slight southward drift, followed by a slight northward one;
    \item over the next 2 days, affected by the second cyclone that generated sustained wind of maximum speed $\approx$15\,m\,s$^{-1}$ over a period of 7\,h at the buoys location, the drift was mainly westward with a slight northward component;
    \item over the last 4--5 days, affected by the third and fourth cyclone that generated winds of speed $\approx$10\,m\,s$^{-1}$, the drift was eastward, first slight southward and then slight northward, similarly to the first phase.
\end{enumerate} 
The phases are divided by sharp turning and looping, hence undergoing significant meandering \citep{gimbert2012tc}.
In total, buoy~B1 drifted 262\,km, mainly zonally ($\approx70$\,km for each of three phases), and exhibits a net northward translation ($\approx80$\,km).
Over the 8 days and 18\,h the average speed was 0.35\,m\,s$^{-1}$ which is over 50\% greater than previously reported daily averages for this sector of the Southern Ocean \citep{heil1999jgr}. The maximum instantaneous speed was 0.75\,m\,s$^{-1}$, which is the fastest recorded for Antarctic pancake ice drift.

Fig.~\ref{f4} shows the development of an ice-edge feature over time, in the form of a localised protrusion that complicates the interpretation of the distance from the edge shown in Fig.~\ref{f3}.
We note, however, that the buoys are always in 100\% ice concentration according to remotely sensed AMSR2 ice concentration.
On 11th~of~July (panel~d), there are large areas covered by intermediate ice concentration around the protrusion ($0\%<i_c<100\%$), likely due to thermodynamic ice formation, which resulted in the sharp increase in distance between buoy~B1 and ice edge on the 11th~of~July shown in Fig.~\ref{f3}.

\subsection{Ice deformations}

\begin{figure}[htbp]
\centering
\includegraphics[width=1\linewidth]{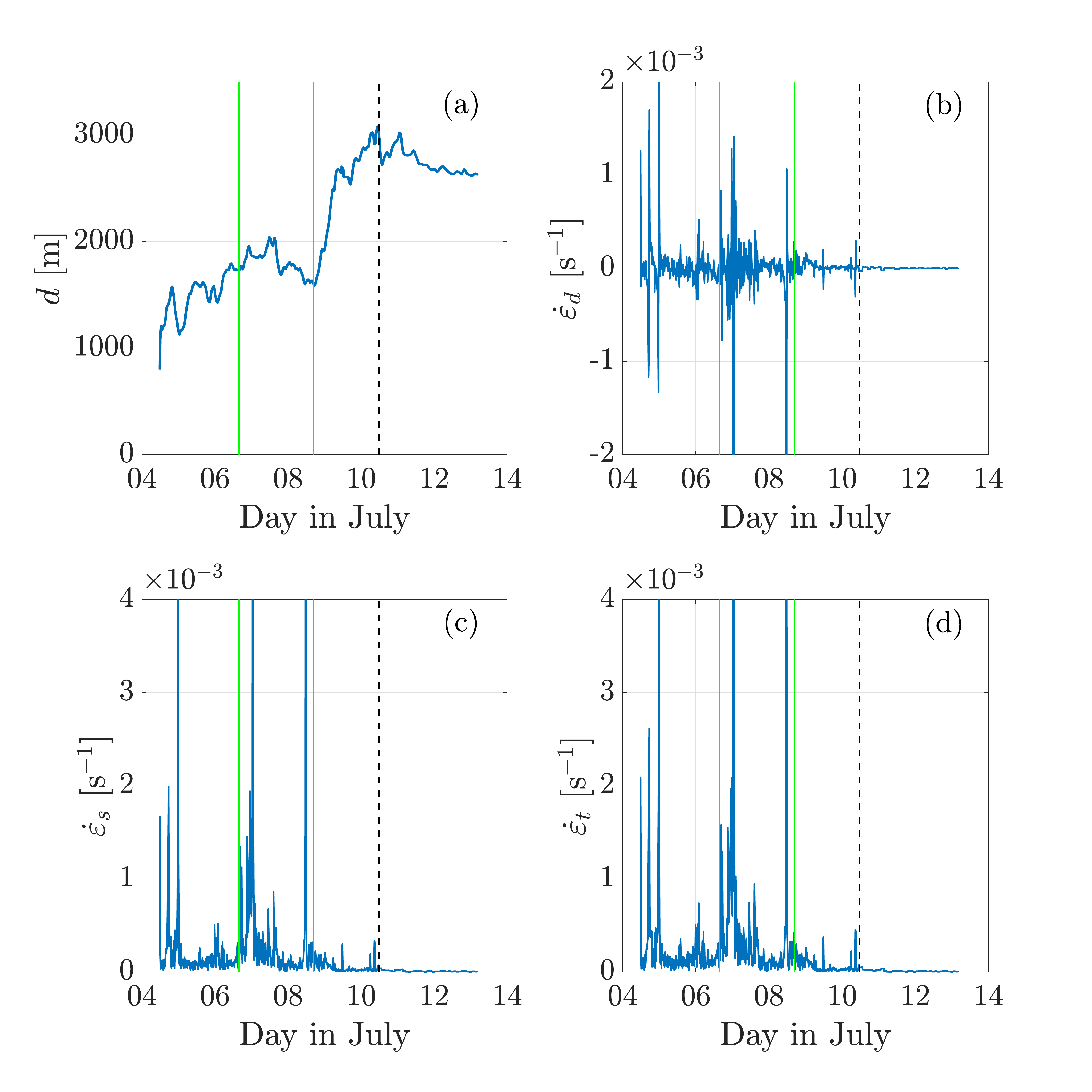}
\caption{(a)~Distance between buoys, (b)~divergence rate, (c)~shear rate, and (d)~total deformation rate. 
The vertical black dashed line denotes the time when the acquisition rate of the second buoy~B2 was dropped from 15\,mins to 2\,h to preserve battery life. The green vertical lines indicate the time of transition between phases.}
\label{f5}
\end{figure}

Fig.~\ref{f5}a shows the time series of the buoy separation distance, $d$. 
Over the first 4 days from deployment (4th--8th July, phases~i--ii), the buoys slowly drifted apart, reaching a maximum separation of $d=$1.5--2\,km.
Over 8th--9th July, at the beginning of phase~(iii), the buoys rapidly drifted apart, from 1.5\,km to 2.5\,km in less than a day, after which they moved slightly closer. 
Overall, the distance between the buoys grew by 2\,km over the nine-day measurement period.

Deformations of the sea ice cover are commonly reported in terms of the strain rates \citep{lindsay2002jgr}
\begin{equation}
\dot{\varepsilon}_d = \frac{\partial u_i}{\partial x} + \frac{\partial v_i}{\partial y}\,,\,\,\,\dot{\varepsilon}_s = \left[\left(\frac{\partial u_i}{\partial x} - \frac{\partial v_i}{\partial y}\right)^2+\left(\frac{\partial u_i}{\partial y} + \frac{\partial v_i}{\partial x}\right)^2\right]^{1/2}\,\textnormal{and}\,\,\,\dot{\varepsilon}_t = (\dot{\varepsilon}_s^2+\dot{\varepsilon}_d^2)^{1/2},
\end{equation}
which are the divergence rate, shear rate  and total deformation rate, respectively. In the equation $u_i$ and $v_i$ denote the ice velocity in the positive east ($x$) and north ($y$) directions, respectively, and the spatial derivatives are evaluated using buoy B1 and B2 velocities ($u_{B1,B2}$ and $v_{B1,B2}$) and position ($x_{B1,B2}$ and $y_{B1,B2}$), e.g. ${\partial u_i}/{\partial x} = {(u_{B1}-u_{B2})}/{(x_{B2}-x_{B1})}$.

Figs.~\ref{f5}b--d show time series of the strain rates.
Divergence and shear intensify at the same time; shear contributes the most to the total strain rate, suggesting that, at the buoy distance length scale, rotational motion dominates over the compression/expansion.
The strain rates are highly intermittent, which provides further evidence that the ice cover remained unconsolidated, at least until the 9th~of~July.
The dashed black vertical lines denote the time at which the B2 sampling rate was lowered to 2\,h, thus reducing the accuracy of the calculations. 
Beyond this time, the calculated deformations are significantly lower and intermittent properties disappear due to the coarse temporal resolution.

The root mean square (RMS) of the strain rates for the time during which both buoys were recording at a sampling rate of 15\,mins are
\begin{equation}
\dot{\varepsilon}_d=7\times 10^{-4}\,\textnormal{s}^{-1}, 
\quad
\dot{\varepsilon}_s=1\times 10^{-3}\,\textnormal{s}^{-1}
\quad\textnormal{and}\quad 
\dot{\varepsilon}_t=1.2\times 10^{-3}\,\textnormal{s}^{-1}.
\end{equation}
These are 2--3 orders of magnitude greater than those reported by \citet{doble2006jgr} for pancake ice in the Weddell Sea, at similar sampling rate of 20\,mins, but from an array of six buoys with characteristic separation distance of 50\,km. 
The large difference is likely due to the disparate characteristic distance between buoys, as rates of deformation are inversely proportional to the distance between buoys \citep{doble2006jgr}, and the strain rates would be comparable if the characteristic distances were equivalent.
Also, computation of the strain rates over the array of six buoys is more accurate than two.

\subsection{Correlation between buoy drift and wind}

\begin{figure}[htbp]
\centering
\includegraphics[width=1\linewidth]{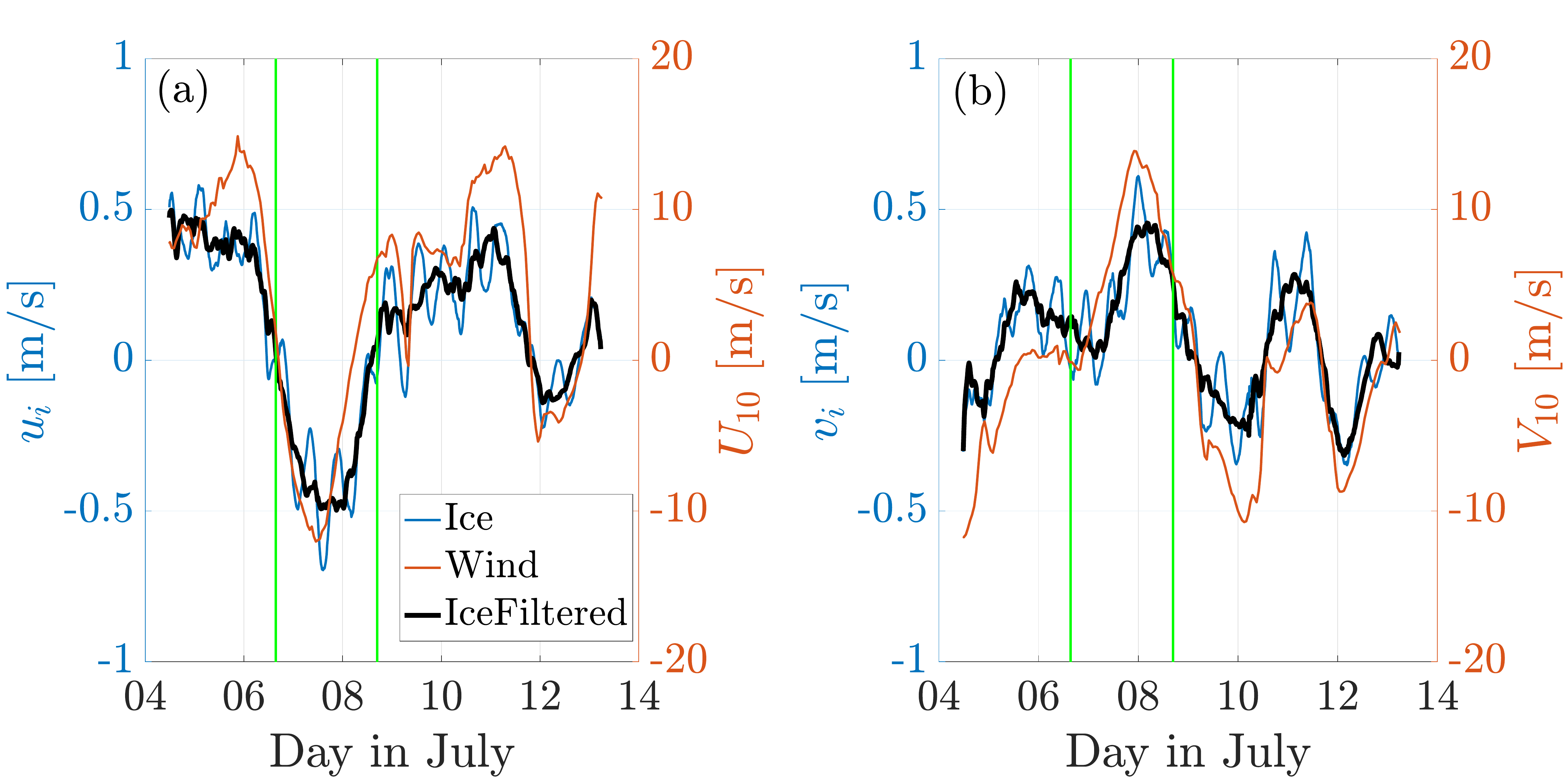}
\caption{(a)~Zonal buoy~B1 velocity (on the left axis) and zonal wind velocity (on the right axis), in black (on the left axis) buoy~B1 zonal velocity when $\approx$13\,h oscillations are excluded. The green vertical lines indicate the time of transition between phases. (b)~As in (a), but for meridional velocity component.}
\label{f6}
\end{figure}

Fig.~\ref{f6} shows the velocity of buoy~B1 in the zonal (east) and meridional (north) directions, compared to the ERA5 wind co-located at buoy~B1 time and position using a tri-linear interpolation (2D in space, 1D in time). The ice drift velocity qualitatively follows the wind velocity, but the ice drift is characterised by oscillations of period $\approx$13\,h.
In Fig.~\ref{f6}a, the wind ($U_{10}$) and ice velocity ($u_i$) are positive (to the east) during phase~(i), and become negative (to the west) during phase~(ii).

\begin{figure}[htbp]
\centering
\includegraphics[width=1\linewidth]{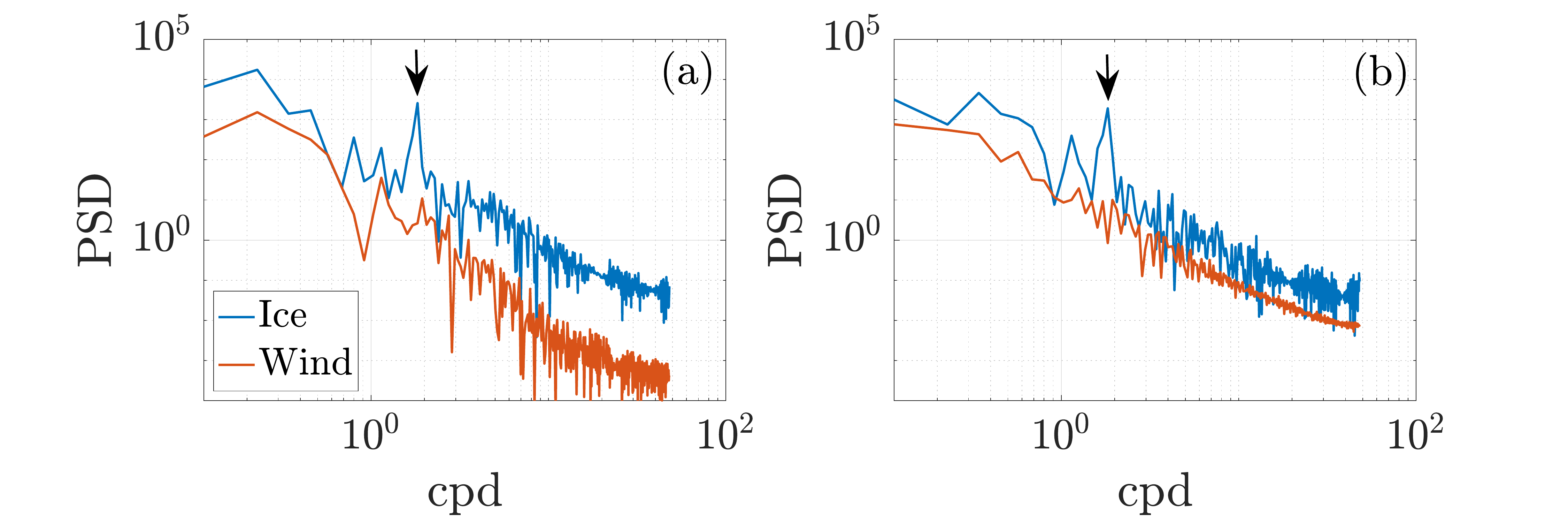}
\caption{(a)~Spectra corresponding to the zonal buoy~B1 velocity (blue) and zonal wind velocity (orange). An arbitrary vertical shift is applied to the wind spectra to aid comparison. Black arrows denote the peak associated to inertial-like oscillations ($\approx$13\,h). (b)~As in (a), but for meridional velocity component.}
\label{f7}
\end{figure}

Fig.~\ref{f7} shows the spectra of the wind velocity components (orange) and ice drift velocity (blue). The wind velocity spectrum forms a continuous energy cascade, but the ice velocity spectrum exhibits an energy peak (highlighted by the arrow) at a frequency just below two cycles per day (cpd; the exact value is 13.1$\pm$0.85\,h). The period of these oscillations is close to the inertial range at 62--63$^\circ$S (13.5$\pm$0.05\,h defined by the Earth rotation) that are clearly seen in the east and north ice velocity components (see Fig.~\ref{f6}).

\begin{figure}[htbp]
\centering
\includegraphics[width=1\linewidth]{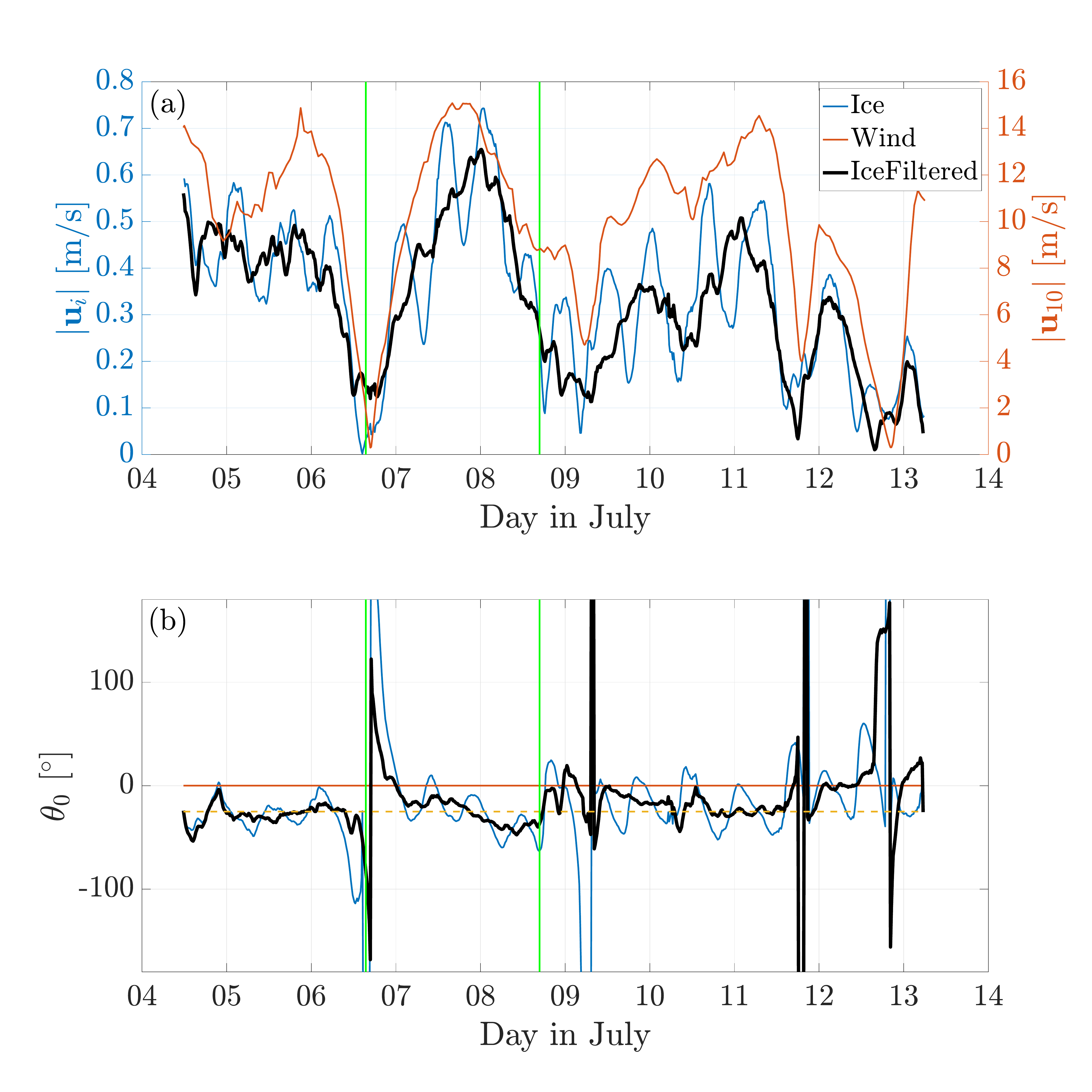}
\caption{(a)~Buoy~B1 speed (left axis) in total (blue) and with $\approx$13\,h oscillations are excluded (black), and total wind speed (right axis; orange). (b)~Difference between ice and wind direction $\theta_0$ (blue), difference when $\approx$13\,h oscillations are excluded (black), and the mean difference $-25^\circ$ (yellow dashed). The green vertical lines indicate the time of transition between phases.}
\label{f8}
\end{figure}

Fig.~\ref{f8}a shows buoy~B1 speed compared to the ERA5 wind speed. 
The instantaneous drift speed peak of 0.75\,m\,s$^{-1}$ occurs during phase~(ii), at midnight between 7th--8th~of~July. 
The wind speed at that instant is 14\,m\,s$^{-1}$, noting that the peak wind speed of 15\,m\,s$^{-1}$ occurs at 16:00 on the 7th~of~July.
During the transition between phase~(i) and phase~(ii), denoted by first vertical line in Fig.~\ref{f8}, the wind stops,~i.e. both the north and east component of the wind velocity approach 0\,m\,s$^{-1}$ (see Fig.~\ref{f6}), and the ice drift almost stops.
The correlation between the wind and buoy~B1 speed is $R^2=0.56$; this increases to $R^2=0.66$ when inertial-like oscillations are filtered out (black curves in Fig.~\ref{f6} and Fig.~\ref{f8}). 
The wind factor, i.e.~the ratio between ice speed and wind speed, estimated with a standard least square regression is 3.3\%. For comparison, \citet{doble2006jgr} report wind factor 3--3.5\% with $R^2=0.5$ in pancake ice.
Little to no correlation is found between the ice drift and the wave-in-ice activity ($R^2<0.1$), even during the periods of large significant wave heights ($H_S>1.25$\,m), suggesting that wave-induced drift of pancake floes is negligible in comparison to wind-induced drift.

Fig.~\ref{f8}\,b shows the angle $\theta_0$ between the ice drift and the wind direction. 
Previous observations indicate the angle, on average, to be in the range 0$^\circ$--30$^\circ$, positive in the northern hemisphere and negative in the southern hemisphere \citep{lepparanta2011book}. 
The present dataset gives a mean angle $\approx -25^\circ$, as shown by the dashed line, and with generally large variations from -$60^\circ$ to $10^\circ$.
The large variations are consistent with, for example, \citet{lund2018jgr}, who reported turning angles from -23$^\circ$ to +83$^\circ$ during a storm event in the Arctic Basin, and in other instances reported ice drift against the wind (turning angle $>90^\circ$).
During our measurements, the largest angles between wind and ice direction ($|\theta_0|>90^\circ$) occur sporadically and are always observed for low wind speed ($|\mathbf{u}_{10}|<6$\,m\,s$^{-1}$), when the wind stresses become small.

\section{Pancake ice drift model}
\subsection{Model formulation}

The Arctic Ice Dynamics Joint Experiment (AIDJEX) model for sea-ice drift \citep{coon1974aidjex,feltham2008ar,lepparanta2011book} is
\begin{equation}\label{eq:aidjex}
m_i\frac{d\mathbf{u}_i}{dt}=A_i\mathbf{S}_a+A_i\mathbf{S}_w+m_i\mathbf{S}_c+m_i\mathbf{S}_g+\nabla\cdot\sigma,
\end{equation}
where $m_i$, $A_i$ and $u_i$ are, respectively, the mass, area and velocity of the ice, $S_a$, $S_w$ and $S_c$ and $S_g$ are external stresses due to wind, ocean currents, Coriolis and ocean tilt, respectively,  and $\nabla\cdot\sigma$ is the rheology that defines internal stresses. 
The ice mass is $m_i=\rho_i h_i A_i$, where $\rho_i$ is the ice density and $h_i$ its thickness.

As stated in \S\ref{env}, in--situ observations of the pancake floes concentration during deployment was $\approx$60\%, and the remaining 40\% was interstitial frazil ice \citep{alberello2019cryo}. 
The wave-in-ice activity measured by the buoys during the subsequent days indicates that similar unconsolidated conditions were maintained.
On this basis, the free drift regime is assumed ($\nabla\cdot\sigma=0$), 
as is standard for low ice concentration \citep{hunke2010jofg,herman2012cejp}.
 
A quadratic wind stress is used, of the form
\begin{equation}
\mathbf{S}_a=C_a \rho_a |\mathbf{u}_a'|\mathbf{u}_a' \exp(\mathrm{i} \theta_a),
\end{equation}
where $C_a$ is the wind drag coefficient over ice, $\mathbf{u}_a'$ the velocity difference between the wind and the ice ($\mathbf{u}_a'=\mathbf{u}_a-\mathbf{u}_i$), $\rho_a$ the air density, $\theta_a$ is the angle between the wind direction and the wind-induced stress, and $\mathrm{i}$ is the imaginary unit.
Linear formulations and calibrated exponents have been used yielding to similar accuracy \citep{martinson1990jgr}, 
but only the more common quadratic formulation is discussed.

For consistency, a quadratic ocean drag is adopted \citep{lepparanta2011book}, with
\begin{equation}
\mathbf{S}_w=C_a \rho_w |\mathbf{u}_w'|\mathbf{u}_w'\exp(\mathrm{i} \theta_w).
\end{equation}
where
\begin{equation}
\mathbf{u}_w'=\mathbf{u}_w+\mathbf{u}_g-\mathbf{u}_i
\end{equation}
assuming ocean and geostrophic currents of velocity $\mathbf{u}_w$ and $\mathbf{u}_g$, respectively. 
We note that
in absence of currents, $\mathbf{u}_w=\mathbf{u}_g=0$, the ocean drag would be proportional to the ice speed and act in the opposite direction to the ice drift, i.e.~$\mathbf{u}_w'=-\mathbf{u}_i$, so that it produces damping.

The term $\mathbf{S}_c$ denotes the Coriolis stress; in absence of other external forces, it produces rotation, which is leftward in the southern hemisphere (with respect to the direction of the ice drift). 
The Coriolis stress is expressed as \citep{cushman2011book}
\begin{equation}
\mathbf{S}_c = - \mathrm{i} f \mathbf{u}_i,
\end{equation}
where $f = 2\omega\sin(\psi)$ is the Coriolis parameter, in which $\omega =7.2921 \times 10^{-5}$rad\,s$^{-1}$ denotes the Earth's rotation rate and $\psi$ is latitude.

The stress due to the ocean slope, $\mathbf{S}_g$, is written \citep{lepparanta2011book}
\begin{equation}
\mathbf{S}_g = - \nabla \zeta,
\end{equation}
where $\zeta$ denotes the sea surface height. 
In deep water this term can be expressed as a function of the surface geostrophic current \citep{cushman2011book}, with
\begin{equation}
\mathbf{S}_g = \mathrm{i} f \mathbf{u}_g
\end{equation}
which is similar in form to $\mathbf{S}_c$, but does not depend on the ice velocity.

The AIDJEX model, Eqn.~\ref{eq:aidjex}, becomes
\begin{equation}
\frac{d\mathbf{u}_i}{dt}=\alpha\mathbf{u}_a'|\mathbf{u}_a'|\exp(\mathrm{i}\theta_a)+\beta\mathbf{u}_w'|\mathbf{u}_w'|\exp(\mathrm{i}\theta_w)-
\mathrm{i} f \mathbf{u}_i + \mathrm{i} f \mathbf{u}_g,
\label{eqn:model}
\end{equation}
where
\begin{equation}
    \alpha=\frac{\rho_{a}C_{a}}{\rho_i h_i}
    \quad\textnormal{and}\quad
    \beta=\frac{\rho_{w}C_{w}}{\rho_i h_i}.
\end{equation}
The Nansen number (the ratio between wind and ocean stresses) can be expressed in terms of the coefficients $\alpha$ and $\beta$, as
\begin{equation}
Na=\sqrt{\frac{\rho_{a}C_{a}}{\rho_{w}C_{w}}}=\sqrt{\frac{\alpha}{\beta}},
\end{equation}
which indicates the wind stresses, explicitly accounting for the air and water drag ratio. 
Eqn.~\ref{eqn:model} is equivalent to the free drift model given by \citet{lepparanta2011book}, Eqn.~6.3,  with the advective acceleration conserved to maintain the generality of our formulation.

\subsection{Model setup}

Eqn.~\ref{eqn:model} is numerically solved in a Lagrangian frame of reference to simulate the buoy drift, using a finite difference, time stepping method; at each step (time steps of 60\,s were found to give sufficient convergence) the velocity and displacement are computed, and the buoy advanced in space.

Wind forcing is retrieved from ERA5 and, at each time step, interpolated in space and time onto the simulated buoy position.
No data are available on ocean currents---in general, currents are rarely available in ice-covered regions \citep{nakayama2012jpo}---thus, $\mathbf{u}_w=0$ is set.
The strong signature of the measured drift at periods close to the inertial range is likely related to the geostrophic current or eddies rather than the Coriolis term, as the contribution of $m_i \mathbf{S}_c$ is negligible for the thin sea ice in the Antarctic marginal ice zone \citep{martinson1990jgr}, and only becomes relevant for multi-year ice ($h_i>1$\,m).

Based on the buoy drift measurements, we adopt the geostrophic term to be a rotational term of the type
\begin{equation}
\mathbf{u}_g = U_g\exp (\mathrm{i} f t),
\end{equation}
where the amplitude of the near-inertial oscillations $U_g$ is estimated from the measurements to be $U_g=0.125$\,m\,s$^{-1}$, which is the mean amplitude of the oscillations in 12--14\,h identified utilising a band-pass filter.
We set $\theta_a=0$ and $\theta_w=-25^\circ$ in agreement with the AIDJEX formulation when surface winds are used \citep{lepparanta2011book}, noting that $\theta_w\approx\theta_0$ \citep{lepparanta2011book}.

The remaining free parameters, $\alpha$ and $\beta$, are calibrated by matching model velocity outputs with the measurements, by minimising the difference $|u_i^O-u_i^M|+|v_i^O-v_i^M|$, where superscripts $O$
and $M$ denote the observations (measurements) and the model, respectively.
We test values of  $\alpha$ in the range 0.012--0.015$\times10^{-3}$\,m$^{-1}$, which corresponds to $C_a$ in 3.0--3.7$\times10^{-3}$, as identified by \citet{overland1985jgr} for the marginal ice zone, and $\rho_i=910$\,kg m$^{-3}$, $\rho_a=1.3$\,kg m$^{-3}$ and $h_i=0.35$\,m, from visual observations during in--situ operations.
Similarly, we test values of $\beta$ in the range 5--16$\times10^{-3}$\,m$^{-1}$, which corresponds to $C_w$ in 1.6--5.0$\times10^{-3}$ and $\rho_w=1028$\,kg m$^{-3}$. 
The lower and upper limits for $C_w$ are taken from values reported  by \citet{martinson1990jgr} in the Weddel Sea, and \citet{mcphee1982techrep} in the Beaufort Sea, respectively.
The calibrated parameters are $\alpha=0.0128\times10^{-3}$\,m$^{-1}$ and  $\beta=8.9\times10^{-3}$\,m$^{-1}$.

\subsection{Model results}\label{sec:modelresults}

\begin{figure}[htbp]
\centering
\includegraphics[width=1\linewidth]{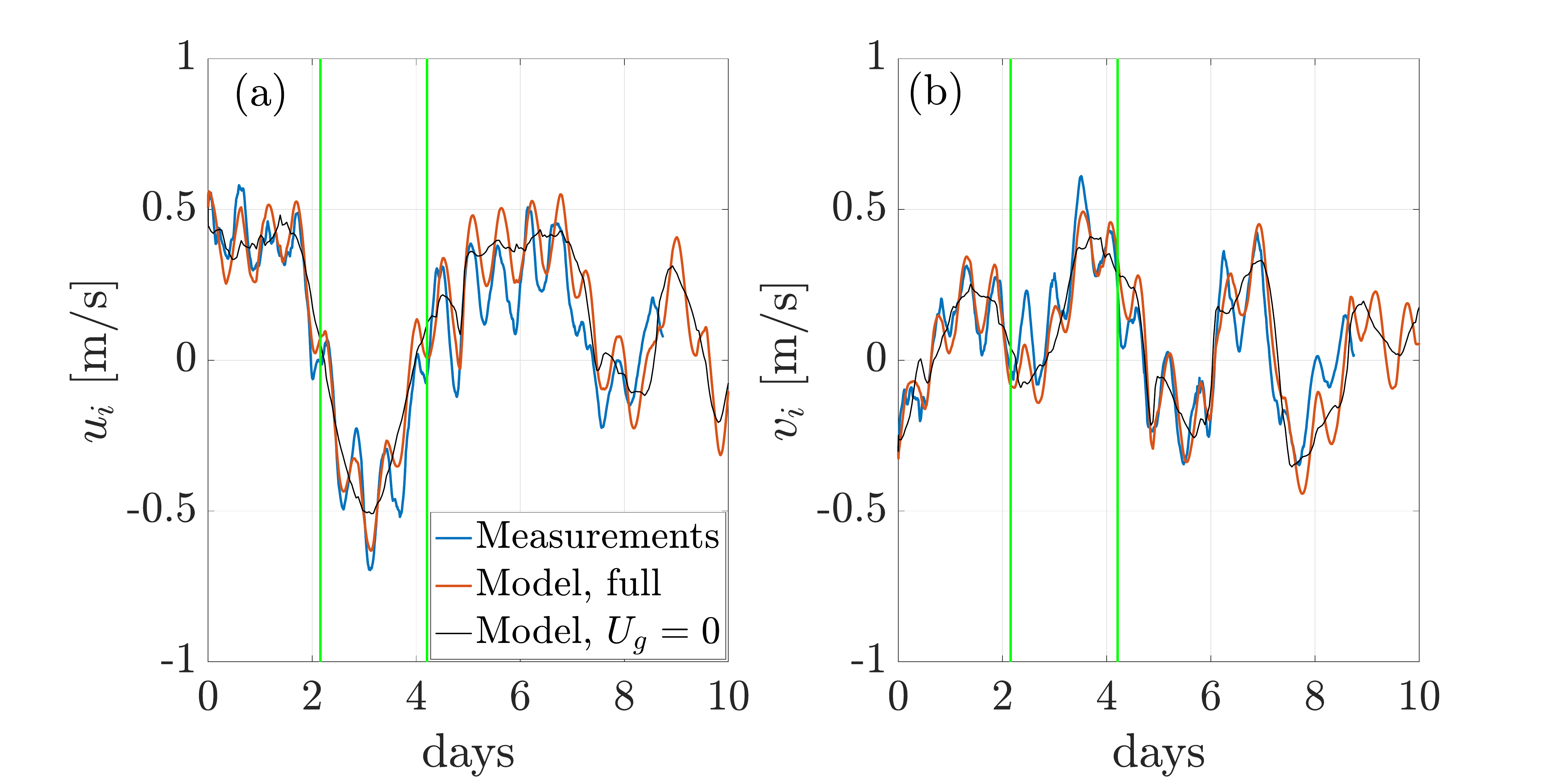}
\caption{Time series of buoy~B1 measurements (blue) and model simulations, for full model (orange) and $U_g=0$ (black): (a)~zonal velocity and (b)~meridional velocity. The green vertical lines indicate the time of transition between phases.}
\label{f9}
\end{figure}

Fig.~\ref{f9} shows model results against measurements for the zonal (a) and meridional (b) velocity components.
Model outputs are shown for both $U_g=0.125$\,m\,s$^{-1}$ (full model) and $U_g=0$, to highlight the effect of the geostrophic term.
Suppression of the geostrophic term ($U_g=0$) eliminates near-inertial, 13\,h-period oscillations, and the time-series resembles the band-pass filtered measurements (black line in Fig.~\ref{f6}).
Model predictions when the geostrophic term is included reproduce the measurements, although some of the high frequency oscillations observed in the measurements are not captured, likely due to relatively low temporal and spatial resolution of the input ERA5 wind data, which results in a smooth wind field, without small scale variability.
The root mean square error over the entire duration of the measurements is 0.095\,m\,s$^{-1}$ for the full model and grows to 0.125\,m\,s$^{-1}$ by suppressing the geostrophic term.

\begin{figure}[htbp]
\centering
\includegraphics[width=1\linewidth]{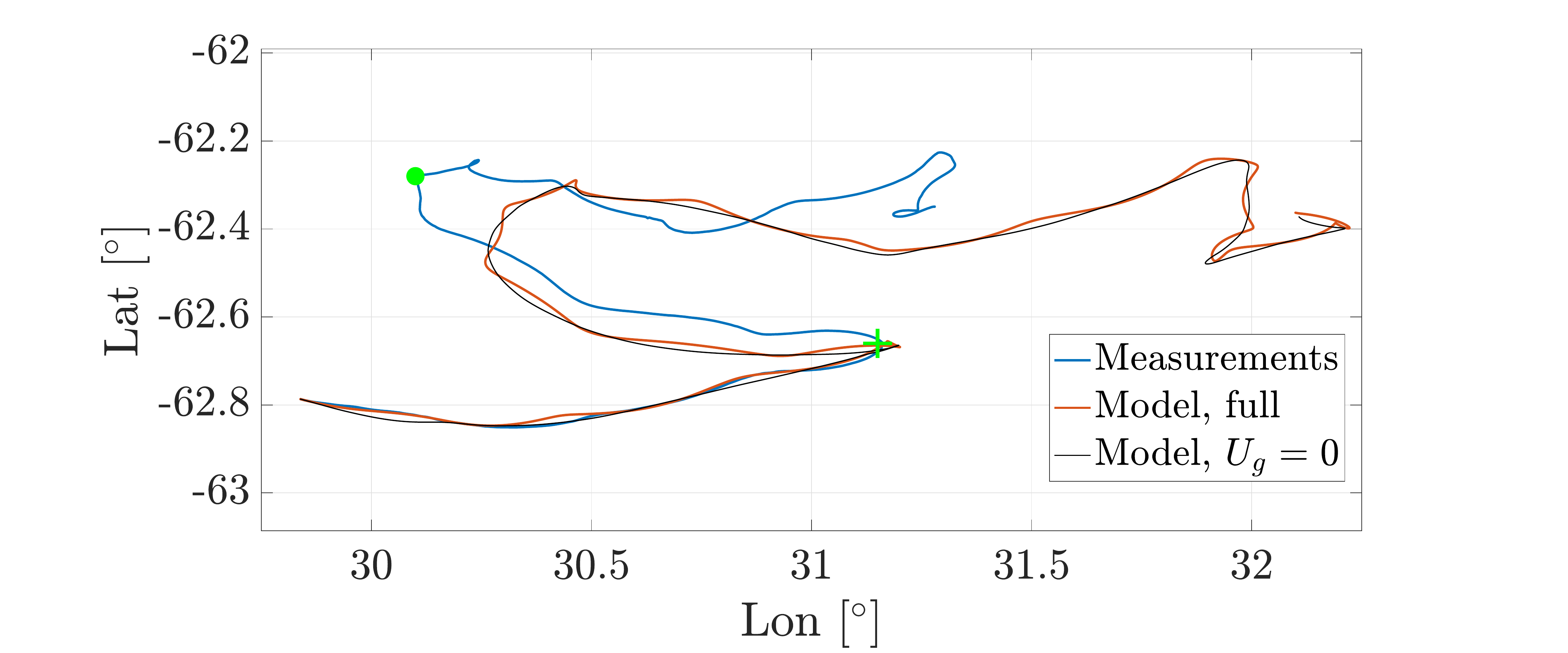}
\caption{Buoy~B1 tracks from measurements (blue) compared to model simulations (full model, orange; $U_{g}=0$, black). The green cross and the green dot indicate the time of transition between phases.}
\label{f10}
\end{figure}

Fig.~\ref{f10} shows the measured and simulated buoy tracks. The full model accurately reproduces buoy~B1 drift during phase~(i), in which the drift is eastward.
After a loop at the end of phase~(i), i.e.~at the location denoted by a green cross in Fig.~\ref{f10}, the model under-predicts the maximum westward movement remaining to the east of the measured buoy position during phases~(ii) and (iii).
Meanders, cycloids (the half-moon shaped part of the track connected by cusps) and loops (during which the rotational component of the motion, driven by the geostrophic forcing, dominates over momentarily weak wind drag) of buoy~B1 track are qualitatively reproduced only when the geostrophic current is included.

\begin{figure}[htbp]
\centering
\includegraphics[width=1\linewidth]{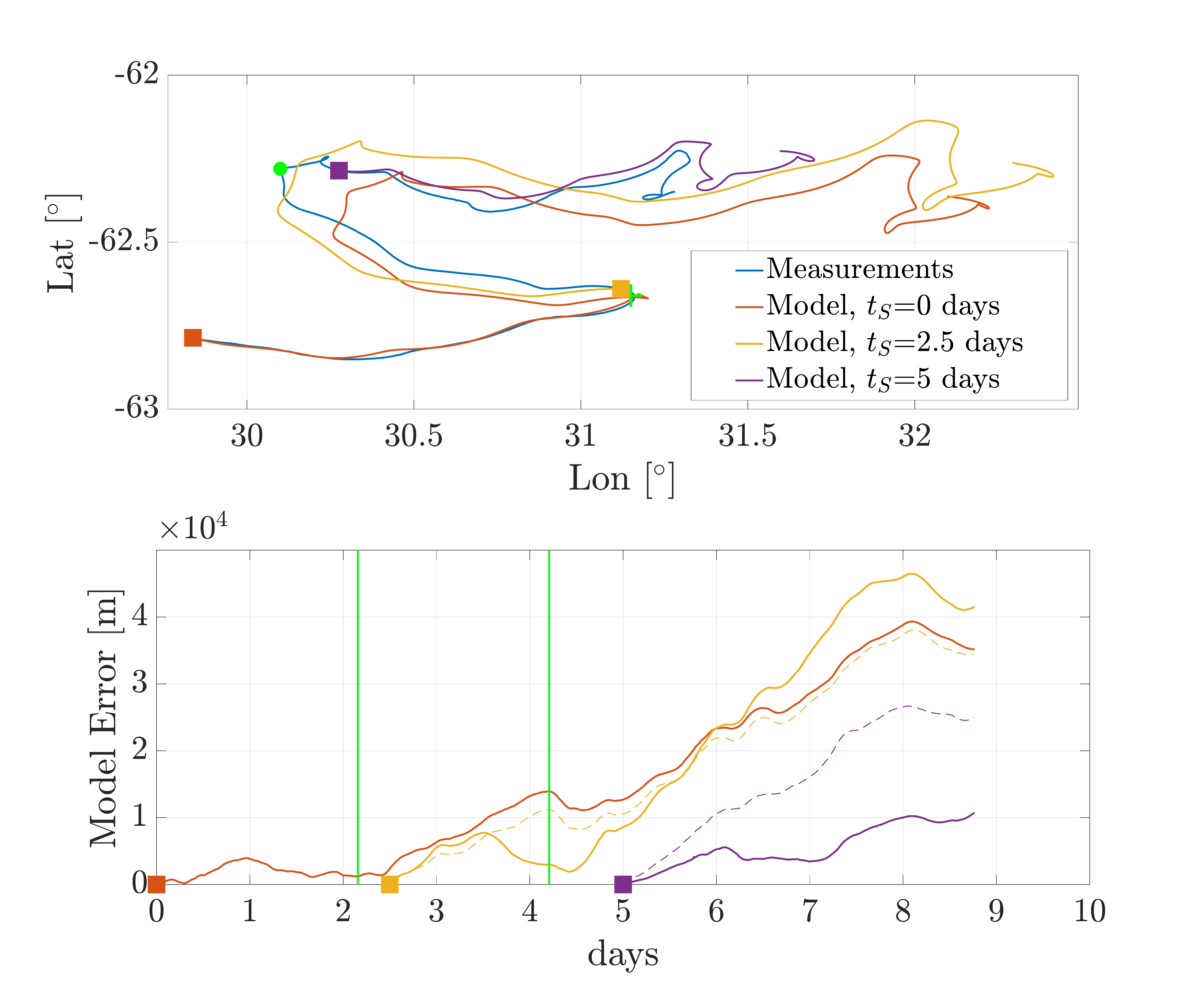}
\caption{(a)~As in Fig~\ref{f10}, but for three different start times ($t_S$, indicated by filled squares), and corresponding tuned parameters $\alpha$ and $\beta$. The green cross and the green dot indicate the time of transition between phases. (b)~Time series of errors for model calibrated for each new start time (solid curves), compared to errors when overall calibrated parameters are used (dashed curves). The green vertical lines indicate the time of transition between phases.}
\label{f11}
\end{figure}

Fig.~\ref{f11}a shows model results that start at $t_S=0$, 2.5\,days and 5\,days from the start of measurements,
noting that $t_S=2.5$\,days is $\approx6$ hrs into phase~(ii) and $t_S=5$\,days is $\approx18$ hrs into phase~(iii).
For the three different start times, parameters $\alpha$ and $\beta$ are calibrated 
over: $t_S=0$ till the end of phase~(i); $t_S=2.5$\,days till the end of phase~(ii); and $t_S=5$\,days till the end of phase~(iii).
Parameters $\alpha$ and $\beta$ are given in Table~\ref{tabula}, noting that the parameters for $t_{S}=0$ (phase~i) are almost identical to the ones calibrated over the entire track.
Compared to phase~(i), the coefficients $\alpha$ and $\beta$ increase during phase~(ii) and decrease during phase~(iii), noting that $\alpha$, which is related to the ice surface roughness \citep{johannessen1983jgr}, is the parameter with the highest variability ($>\pm50\%$). 

\begin{table}
\caption{Calibrated parameters for the start times $t_{s}=0$, 2.5\,days and 5\,days (roughly phases~i--iii, respectively). The values in parenthesis for phases~(ii--iiii) show the variation compared to phase~(i).} 
\centering
\begin{tabular}{l l l l}
\hline
\textbf{} & $\alpha\times10^{-3}$\,[m$^{-1}$] & $\beta\times10^{-3}$\,[m$^{-1}$] & $Na\times10^{-2}$ [-] \\
\hline
   Phase~(i) & 0.0128 (--) & 8.9 (--) & 3.81 (--) \\
   Phase~(ii) & 0.0225 ($+75.8\%$) & 11.6 ($+30.3\%$) & 4.40 ($+15.5\%$)\\
   Phase~(iii) & 0.0061 ($-52.3\%$)& 7.4 ($-16.9\%$) & 2.87 ($-24.7\%$)\\
\hline
\end{tabular}
\label{tabula}
\end{table}

Fig.~\ref{f11}b shows the time series of model errors corresponding to Fig.~\ref{f11}a, 
i.e.~distances between the model and measured positions, for the start times $t_S=0$, 2.5\,days and 5\,days. 
It also includes errors for the start times $t_S=2.5$\,days and 5\,days, without re-calibration of $\alpha$ and $\beta$.
The error for $t_{S}=0$\,days ($\alpha=0.0128\times10^{-3}$\,m$^{-1}$ and $\beta=8.9\times10^{-3}$\,m$^{-1}$) is $<5$\,km during phase~(i) and only exceed this threshold at 2.75\,days.
The error then steadily grows (during phases~ii---iii), up to $\approx$40\,km at 8\,days,
due to error propagation in the time integration, and changes in the optimal values of $\alpha$ and $\beta$. 
The error for $t_{S}=2.5$\,days ($\alpha=0.0225\times10^{-3}$\,m$^{-1}$ and $\beta=11.6\times10^{-3}$\,m$^{-1}$) never exceeds 7.5\,km during phase~(ii) and at the end of phase~(ii) the model error is $<3$\,km. For comparison, utilising parameters calibrated over the entire track the error at the end of phase~(ii) is $\approx$11\,km, i.e.~4 times larger than the model error with dedicated parameters.
The error for $t_{S}=5$\,days ($\alpha=0.0061\times10^{-3}$\,m$^{-1}$ and $\beta=7.4\times10^{-3}$\,m$^{-1}$) is only 3.4\,km after 2\,days, and remains $<11$\,km till the end of phase~(iii). 
This is significantly better than model prediction for $t_{S}=5$\,days and parameters calibrated over the entire track, 
which, for example, result in an error of 27\,km after 3\,days.

\section{Discussion}\label{sec:discuss}

Pancake ice constitutes most the winter ice mass budget around Antarctica and it is becoming more common in the emerging Arctic marginal ice zone \citep{wadhams2018jgr}. 
Thermodynamics and dynamics of pancake ice govern the atmosphere--ocean--sea ice momentum and mass exchanges over vast ice covered areas, thus playing a role in the global climate system \citep{doble2003pancake,smith2018jgr}.

This study is the first to measure and analyse both the drift of pancake ice floes and concomitant wave activity, during a series of intense winter polar cyclones. 
The analysis is based on the assumption that pancake ice conditions persisted over the nine days following deployment.
Cyclonic activity and associated intense wave-in-ice activity prevents consolidation of pancake ice floes \citep{shen2004jgr,doble2006jgr}. 
Their absence has been used to infer consolidation of the pancakes into a compact ice cover \citep{doble2006jgr}.
Our measurements of energetic waves ($H_S>1.25$\,m) and intermittent internal sea ice deformations 100--200\,km from the ice edge suggest that pancake ice conditions similar to the ones at deployment \citep{alberello2019cryo} persisted for at least the first 7 days following deployment, beyond which ice conditions may have transformed as the ice edge advanced.

Despite the significant wave-in-ice activity, no evidence was found of wave-induced ice drift, as might be caused by  Stokes drift \citep{yiew2017pof}, slope-sliding \citep{grotmaack2006jwpcoe} or wave radiation stresses \citep{masson1991jpo}.
\citet{williams2017tc} and \citet{boutin2019cryo} recently integrated wave radiation stresses  into large-scale numerical models that include wave attenuation and wave-induced ice breakup, based on the wave--ice interaction model of \citet{williams2013oma,williams2013omb}.
They found that large wave radiation stresses, proportional to the wave attenuation rate, remain concentrated at the edge \citep{williams2017tc}; wind and ocean stresses dominate ice drift over longer distances. 
Moreover, \citet{williams2017tc} found that wave-radiation stresses are appreciable only for wave periods $<10$\,s; the measurements reported here have dominant periods $>15$\,s, and also for smaller floes than tested by \citet{williams2017tc}, for which radiation stresses are even weaker.
Although wave-induced drift is negligible, it is expected that the significant waves measured will have induced turbulence in the water sublayers \citep{zippel2016elementa,alberello2019wm,smith2019jgr}, enhancing mixing and heat fluxes under sea ice \citep{ackley2015aog,smith2018jgr}.

\citet{shen1987jgr} proposed a granular rheology for the marginal ice zone based on momentum transfer through floe--floe collisions. 
\citet{feltham2005ptra} used the collisional rheology in a compositive marginal ice zone/pack ice rheology, and \citet{bateson2019cryo} included wave stresses in the rheology.
In comparison, \citet{sutherland2018jpo} used a rheology  based on Mohr--Coulomb granular theory, and their model outputs and field measurements showed strong wave attenuation and ice deformation that resulted in rafting of the floes.
Notably, the ice drift was constrained by the coast, allowing for the internal stresses to build up \citep{dai2004jgr}.
The model--data agreement shown in \S\ref{sec:modelresults}, without a rheology term, indicates  internal stresses are negligible for pancake ice during intense cyclones conditions, and no collisions or rafting were observed during deployment. 
This is consistent with laboratory wave basin experiments reported by \citet{bennetts2015ptrsa}, which showed negligible attenuation, and although regular floe--floe collisions occurred, they were weak and did not result in rafting. 
Discrete element models of pancake ice floes in waves \citep{hopkins2001aog,sun2012crst} also show that no rafting occurs in open boundary configuration, but it does when waves push the floes against a fixed boundary \citep{dai2004jgr}.

Drift measurements conducted at high temporal resolution generated accurate estimates of the drift speed \citep{thorndike1986} under cyclonic activity.
The speed reached $\approx$0.75\,m\,s$^{-1}$, which is the highest ice speed ever recorded in the Southern Ocean.
Evaluation of the drift speed is sensitive to the sampling rate \citep{thorndike1986} and daily or sub-daily measurements, available using remote sensing products (e.g.~OSI-SAF; \citet{lavergne2010jgr}), can underestimate the maximum ice drift speed by over 20\%, making them unsuitable to study drift at small temporal scales.
A detailed analysis of our data indicates that the maximum speed is reduced by $\approx5$\% when the sampling is lowered to 6\,h, and by $\approx$20\% for 12\,h sampling.   
Velocity components in the north and east directions show larger reductions.
Previously reported measurements at a 6\,h sampling rate or greater \citep{martinson1990jgr,vihma1996jgr,heil1999jgr} might have underestimated the instantaneous drift speed and, consequently, provided lower estimates of the drag coefficients and wind factors over sea ice.

Low temporal resolution drift measurements would not have captured the oscillations with period close to the inertial range; at least a 3\,h resolution is needed to capture these oscillations.
The rotational motion significantly contributes to instantaneous ice speed, and induces instantaneous ice drift in opposition to the wind direction, when the wind intensity drops.
The model outputs indicate Coriolis forcing is not responsible for the observed oscillations, i.e.~model results including Coriolis forcing and omitting geostrophic forcing do not reproduce the observed velocity oscillations, even for thicker ice, up to 1\,m. 
Moreover, tidal currents have previously been found to affect ice drift only in limited water depth conditions \citep{meyer2017jgr,peterson2017jgr,padman2018jgr}, especially in shelf seas and coastal areas, and, therefore, are unlikely to be the source of the periodic oscillations since the study area is located in deep waters \citep{arndt2013grl}.
Instead, combined measurements and model outputs support the existence of a geostrophic-like forcing at period close to 13\,h, similarly to the indirect observations of \citet{lund2018jgr} in the Arctic.
The  rotational motion period and amplitude ($\approx$2\,km in diameter) are consistent with sub-mesoscale eddies that have been found to form at the edge of the marginal ice zone in the Arctic \citep{lund2018jgr} and in numerical experiments \citep{manucharyan2017jgr,dai2019om}.

Our analysis indicates that the ratio $C_a/C_w$, using a quadratic drag formulation, is order unity for pancake ice in the Southern Ocean winter marginal ice zone---this value is obtained using $\rho_i=910$\,kg m$^{-3}$, $\rho_a=1.3$\,kg m$^{-3}$, $\rho_w=1028$\,kg m$^{-3}$ and $h_i=0.35$\,m estimated at deployment, which gives $C_a = 0.0032$, $C_w = 0.0027$. 
The values for the drag coefficients are close to those found in the marginal ice zone by \citet{overland1985jgr}, \citet{martinson1990jgr}, \citet{mcphee1982techrep} and \citet{lepparanta2011book},
noting that none of previously reported drag coefficients explicitly refers to pancake ice.
Moreover, sea ice drag coefficients do not account for roughness due to ocean waves propagating in the marginal ice zone, which remains an open problem \citep{zippel2016elementa}.

\citet{lepparanta2011book} argues that the ratio $C_a/C_w$ does not vary for all ice types because the ice roughness on the air and water sides are correlated, and hence the Nansen number $Na=\sqrt{\alpha/\beta}\propto \sqrt{C_a/C_w}$ does not vary (assuming air and water densities are constant).
Model calibrations of the parameters $\alpha$ and $\beta$ indicate variation over the nine days following deployment: $Na=0.0381$ in phase~(i); $Na=0.0440$ in phase~(ii); and $Na=0.0287$ in phase~(iii).
Evolution of the Nansen number indicates a corresponding change of the ratio $C_a/C_w$, 
suggesting the ice conditions modified two days after deployment, i.e.~at the transition from phase~(i) to phase~(ii), and again after four days, i.e.~at the transition from phase~(ii) to phase~(iii), noting that phase~(iii) is characterised by less intense wave-in-ice activity and significantly slower drift.
However, model calibrations crucially depend on the input wind and currents, and we recall that no data on currents are available for our experiments, and a thorough analysis of ERA5 wind bias in ice covered region is needed.

\section{Conclusions}

High temporal resolution measurements of drift of a pair of buoys deployed on pancake floes, initially 100\,km into the marginal ice zone, during the Antarctic winter expansion were analysed over a 9-day period, over which four polar cyclones impacted the ice cover.
The measurements, and comparisons with a calibrated Lagrangian free drift model, revealed that:
\begin{itemize}
    \item{Pancake ice floes in the marginal ice zone are extremely mobile, even in 100\%  ice concentration (60\% pancake ice and 40\% interstitial frazil ice). The maximum instantaneous ice drift speed was 0.75\,m\,s$^{-1}$, measured during intense storm conditions (winds up to $\approx$15\,m\,s$^{-1}$), and exceeding previously reported values for the ice-covered Southern Ocean.}
    \item{Pancake ice drift velocity correlates very well with wind velocity, indicating that wind is the dominant forcing, except for a strong inertial-like signature at $\approx$13\,h in the drift, which was attributed to geostrophic currents (or sub-mesoscale eddies). Despite the strong wave-in-ice activity, no correlation was found with the measured ice drift.}
    \item{A free drift model accurately predicts pancake ice drift velocities, indicating that internal stresses are negligible. This finding was backed by the relative motion between the buoys, which was two orders of magnitude smaller than the total drift.}
    \item{The Nansen number varied considerably over the nine days period at the scale of synoptic events (2--3 days)
    suggesting that ice conditions and, consequently, ocean and wind drag have changed,
    although it may also be due to inaccuracies in the model forcings used.}
\end{itemize}    

Present results highlight the need for better understanding and models of $\alpha$ and $\beta$ (equivalently, the drag coefficients and ice thickness) and their temporal and spatial variation, together with reliable wind and current data. 
This will empower accurate predictions of pancake ice drift in the marginal ice zone at the 2--3~day temporal scale of synoptic events, particularly during polar cyclones continuously reshape the marginal ice zone and have the largest effect on the advance and retreat of pancake ice around Antarctica.

\section*{Acknowledgments}

The expedition was funded by the South African National Antarctic Programme through the National Research Foundation.
This work was motivated by the Antarctic Circumnavigation Expedition (ACE) and partially funded by the ACE Foundation and Ferring Pharmaceuticals.
AA, LB, AT and PH were supported by the Australian Antarctic Science Program (all by project 4434, and PH by projects 4301 and 4390).
MO was supported by the Departments of Excellence 2018–2022 Grant awarded by the Italian Ministry of Education, University and Research (MIUR) (L.232/2016).
CE was supported under NYUAD Center for global Sea Level Change project G1204.
PH was supported by the Australian Government’s Cooperative Research Centres Programme through the Antarctic Climate and Ecosystems Cooperative Research Centre.
ERA5 reanalysis was obtained using Copernicus Climate Change Service Information 2019.
AA and AT acknowledge support from the Air-Sea-Ice-Lab Project.
MO acknowledges B GiuliNico for interesting discussions.

\bibliographystyle{agsm}
\bibliography{biblio}  


\end{document}